\documentclass[aps,prb,twocolumn,groupedaddress,floatfix,showpacs,amsmath,amssymb]{revtex4}
\usepackage{graphicx}
\usepackage{amsmath}

\begin{document}
\newcommand\ket[1]{|#1\rangle}
\newcommand\bra[1]{\langle#1|}
\newcommand\braket[2]{\left\langle#1\left|#2\right.\right\rangle}

\title{Pumping through a quantum dot in the proximity of a superconductor}

\author{Janine Splettstoesser}
\affiliation{NEST-CNR-INFM \& Scuola Normale Superiore, I-56126 Pisa, Italy}
\affiliation{Institut f\"ur Theoretische Physik III,
Ruhr-Universit\"at Bochum, D-44780 Bochum, Germany}
\affiliation{D\'epartement de Physique Th\'eorique, Universit\'e de Gene\`eve,
CH-1211 Gen\`eve 4, Switzerland}
\author{Michele Governale}
\affiliation{Institut f\"ur Theoretische Physik III,
Ruhr-Universit\"at Bochum, D-44780 Bochum, Germany}
\author{Fabio Taddei}
\affiliation{NEST-CNR-INFM \& Scuola Normale Superiore, I-56126 Pisa, Italy}
\author{J\"urgen K\"onig}
\affiliation{Institut f\"ur Theoretische Physik III,
Ruhr-Universit\"at Bochum, D-44780 Bochum, Germany}
\author{Rosario Fazio}
\affiliation{International School for Advanced Studies (SISSA), via Beirut 2-4, I-34014 Trieste, Italy}
\affiliation{NEST-CNR-INFM \& Scuola Normale Superiore, I-56126 Pisa, Italy}
\date{\today}
\begin{abstract}

We study adiabatic pumping through a quantum dot tunnel-coupled to one normal and one superconducting lead. We generalize a formula which relates the pumped charge through a quantum dot with Coulomb interaction to the instantaneous local Green's function of the dot, to systems containing a superconducting lead. First, we apply this formula to the case of a non-interacting, single-level quantum dot in different temperature regimes and for different parameter choices, and we compare the results with the case of a system comprising only normal leads. Then we study the infinite-U Anderson model with a superconducting lead at zero temperature, and we discuss the effect of the proximity of the superconductor on the pumped charge.

\end{abstract}

\maketitle

\section{Introduction}
A finite charge can be pumped through a mesoscopic system in
the absence of an applied bias voltage by changing
periodically in time some parameters of the system. If the
parameters are changed in time slowly with respect to the lifetime
of the electrons in the system, pumping is \textit{adiabatic}. The
idea of electron pumping through a mesoscopic system is based on a
work of Thouless.\cite{thouless83} Since then a large number of
theoretical
\cite{brouwer98,makhlin01,buttiker01,buttiker02,zhou99,entin02,avron00,aunola03,
mottonen06,wang01,blaauboer02,wang02,taddei04,pekola99,fazio03,niskanen03,
splett05,sela06}
and experimental \cite{pothier92,switkes99,geerligs91,fletcher03}
works has been dedicated to this field. Adiabatic pumping has a
geometric nature: The pumped charge depends only on the area spanned by 
the pumping cycle in parameter space and not on its detailed timing. 
It can therefore be related to geometric phases.\cite{avron00,makhlin01,aunola03,mottonen06} 
In normal conductors transport is due to the transfer of
individual quasiparticles. In superconducting devices, quasiparticle transport is hindered by the gap, 
and charge pumping is due to Cooper pairs. 
\cite{mottonen06,pekola99,fazio03,niskanen03} 
In hybrid normal-superconducting systems, Andreev reflection at the interface between the normal and the superconducting region allows for sub-gap charge transport in the system. 
In the absence of Coulomb interaction, adiabatic pumping in such hybrid systems 
can be studied by generalizing in Nambu space the scattering approach to adiabatic pumping.
\cite{wang01,blaauboer02,wang02,taddei04}

\begin{figure}
\includegraphics[width=2.8in]{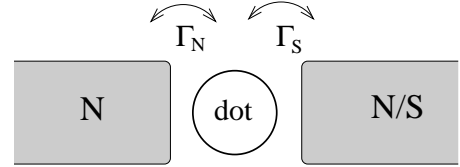}
\caption{Sketch of the dot attached to two leads. We are interested in the N-dot-S system, where the left lead is
normal conducting and the right lead is superconducting and will compare to the N-dot-N system, where both leads
are normal conducting. Left and right lead are attached to the dot via tunnel barriers. The tunneling strength is
described by the respective rate $\Gamma_{\mathrm{N}}$ and $\Gamma_{\mathrm{S}}$. Possible time-dependent parameters are: the strength of the tunnel couplings $\Gamma_{\mathrm{N}}$ and $\Gamma_{\mathrm{S}}$, and the dot-level position $\epsilon$. 
\label{fig_system}}
\end{figure}

In the present paper we study a system consisting of a quantum-dot attached to a normal conducting lead and a superconducting
lead (N-dot-S system) as shown in Fig. \ref{fig_system}. 
In the absence of Coulomb repulsion in the dot, adiabatic pumping through such a system has been discussed in Refs. \onlinecite{wang01,blaauboer02,wang02}. On the other hand, when 
Coulomb repulsion is present, only DC transport through the N-dot-S system has been investigated.\cite{fazio98,kang98,schwab99,clerk00,lambert00,cuevas01}\\

We generalize a
recently developed formalism to calculate the pumped charge through an interacting quantum dot\cite{splett05,sela06} to the case of the N-dot-S system.
First, we apply this formalism to a non-interacting quantum
dot,  studying systematically the ratio between the
charge pumped through the N-dot-S system and the charge pumped
through a system consisting of a dot attached to two normal
conducting leads (N-dot-N system). This
ratio was calculated in Refs.
\onlinecite{wang01,blaauboer02,wang02} for a non-interacting
resonant system at zero temperature. We consider finite
temperature and extend the analysis to the off-resonant situation.
 Finally, we apply the formalism  to the infinite-U
Anderson model in the Kondo regime. We discuss the pumped charge
as a function of the level position and the superconducting gap
and compare to the interacting N-dot-N system.

\section{Model and Formalism}
In this paper we consider a quantum dot tunnel coupled to one normal and one superconducting lead, as shown in Fig. \ref{fig_system}.
The system is described by the Hamiltonian: $H=\sum_{\alpha=\text{N,S}}(H_{\alpha}+H_{\text{tunn},\alpha})+H_{\text{dot}}$. 
The dot Hamiltonian  reads $H_{\text{dot}}=\sum_\sigma
\epsilon (t) d_{\sigma}^\dagger d^{}_\sigma+ U n_{\uparrow} n_{\downarrow}$, 
where $\sigma$ is a spin label, $n_\sigma=d_{\sigma}^\dagger d^{}_\sigma$ and we allow the dot level $\epsilon$ to depend on time $t$. 
The Hamiltonian for the normal lead reads $H_{\text{N}}=\sum_{k,\sigma}
\epsilon_k c_{N,k\sigma}^\dagger c^{}_{N,k\sigma}$, and the one for the superconducting lead reads $H_{\text{S}}=\sum_{k,\sigma}
\epsilon_k c_{S,k\sigma}^\dagger c^{}_{S,k\sigma}+\sum_k (\Delta
c_{S,k\uparrow}^\dagger c_{S,-k\downarrow}^\dagger+H.c)$, with $\Delta\in
\mathbb{R}$ being the superconducting order parameter.
We take $\Delta$ to be
$k$-independent, which is appropriate for conventional BCS superconductors such as aluminum and niobium.
 The operator  $d_{\sigma}^\dagger$ ($d_{\sigma}$) creates (annihilates) a dot electron with spin $\sigma$, the operator $c_{\alpha,k\sigma}^\dagger$  ($c_{\alpha,k\sigma}$) creates (annihilates) an electron in lead $\alpha$ with momentum $k$ and spin $\sigma$.
The tunnel coupling between the dot and the leads is taken into account by the  tunneling Hamiltonians $H_{\text{tunn},\alpha}= \sum_{k,\sigma}(
V_\alpha(t) c_{\alpha,k\sigma}^\dagger d_\sigma+H.c.)$, where $\alpha=\text{N,S}$, and the tunneling 
amplitudes $V_\alpha$ can be time-dependent.
We assume that the phase of $V_\alpha$ does not depend on time, since a time-dependent 
phase would describe an applied bias voltage. 
From now on we work in Nambu space, defining the fields:
$\Psi_{\alpha,k}=(c^{}_{\alpha,k\uparrow},c_{\alpha,-k\downarrow}^\dagger)$, and 
$\Phi=(d^{}_{\uparrow},d_{\downarrow}^\dagger)$.

Generalizing the approach of Refs.~\onlinecite{splett05,sela06}, the pumped current can be related to the 
local Green's function of the  dot. In particular, the first adiabatic
correction to the 
charge current from the dot to the left (normal) conductor through the barrier 
is given by
\begin{widetext}
\begin{equation}
  \label{adiabaticcurrent}
J(t)  =  -\frac{e}{2\pi} \int d\omega \left(- \frac{\partial f}
  {\partial \omega} \right) 
  \mathbb{R}\mathrm{e}\left\{\mbox{Tr} \left[\tau_3
  \frac{d}{d t} \left[ \Gamma_{\text{N}}(t) \hat{G}_{0}^{\text{r}}(\omega,t)
  \right]
  \left( \hat{G}_{0}^{\text{r}}(\omega,t) \right)^{-1}
  \hat{G}_0^{\text{a}}(\omega,t)
  \right]\right\}+J^{\text{corr}}, 
\end{equation}
\end{widetext}
where $f(\omega)$ is the Fermi function, the caret indicates a matrix in Nambu space and $\tau_3$ is the Pauli matrix given by $\tau_3=\left(\begin{array}{cc} 1&0\\0&-1\end{array}\right)$. 
The first term in Eq.~(\ref{adiabaticcurrent}) corresponds to the so called average-time 
approximation,\cite{splett05} while  $J^{\text{corr}}$ (discussed below) contains contributions due to vertex corrections\cite{sela06} and it is zero for all cases studied in this paper. 
The instantaneous retarded Green's function is defined as 
$\hat{G}_{0}^{\text{r}}(\tau,0,t)=-i \Theta(\tau) \langle \{ \Phi^\dagger(\tau),\Phi(0)\}\rangle$, and its Fourier transform is $\hat{G}_{0}^{\text{r}}(\omega,t)= \int d\tau \, e^{i\omega\tau}\hat{G}_{0}^{\text{r}}(\tau,t)$. The last argument ($t$) of the Green's function is the time with respect to which the adiabatic expansion has been performed.\cite{splett05} In particular, the instantaneous retarded Green's function, $\hat{G}_0^{\text{r}}(\omega,t)$ is computed with  all time-dependent parameters frozen at time $t$. The advanced Green's function is related to the retarded one by 
$ \hat{G}_0^{\text{a}}(\omega,t) =
(\hat{G}_0^{\text{r}}(\omega,t))^\dagger$. The intrinsic line width $\Gamma_\alpha$ is defined as $\Gamma_\alpha(t,t')=2\pi \rho_\alpha V^{}_\alpha(t)V^*_\alpha(t')$ with $\alpha=\mathrm{N,S}$ and $\Gamma_\alpha(t)=\Gamma_\alpha(t,t)$. The normal lead is supposed to have flat bands with a constant density of states $\rho_\mathrm{N}$.
The density of states  $\rho_\mathrm{S}$ is the one of the superconducting lead in its normal state and it is constant, too.

We now discuss the average-time approximation and give an expression for the correction term $J^{\text{corr}}$. 
The average time approximation neglects terms due to vertex corrections\cite{sela06} 
in evaluating the adiabatic expansion of the self energy. 
The self energy is defined by the Dyson equation $G\left(t',t'',t\right)=g\left(t',t'',t\right)+\int dt_1\int dt_2 G\left(t',t_1,t\right)\Sigma\left(t_1,t_2,t\right)g\left(t_2,t'',t\right)$ and it is a functional of the time-dependent Hamiltonian $H(\tau)$, where $\tau\in\left[t_1,t_2\right]$. The adiabatic expansion is performed by linearizing the time-dependence of the Hamiltonian around a fixed time $t$ (denoted as a third argument of the self energy). In order to solve the Dyson equation for the adiabatic expansion of the Green's function, the average-time approximation is performed on top of the adiabatic expansion.
Such approximation consists in replacing the linear time dependence on $\tau$ by the average time $(t_1+t_2)/2$ 
of the interval $\left[t_1,t_2\right]$ and thus it neglects the term 
\begin{eqnarray*}
&&\Sigma^{\mathrm{corr}}\left(t_1,t_2,t\right)=\\
&&T_{\mathrm{K}}\int_{\mathrm{K}}d\tau\left.\frac{\delta\Sigma}{\delta H\left(\tau\right)}\right|_{H\left(\tau\right)=H\left(t\right)}\left(\tau-\frac{t_1+t_2}{2}\right)\dot{H}_{\tau}\left(t\right)
\nonumber \ ,
\end{eqnarray*}
where the time integral is calculated along the Keldysh contour, $T_{\mathrm{K}}$ is the time-ordering operator along the Keldysh contour, and $\delta\Sigma/\delta H\left(\tau\right)$ a functional derivative. 
Finally, the correction to the pumped current can be written as
\begin{eqnarray}
&&J^{\text{corr}} =\label{eq_corr}\\ 
&&\frac{e}{2\pi}\int d\omega \int \frac{d\omega'}{\pi} \mathbb{R}\mathrm{e}\mbox{Tr} \left[\hat{G}_0^{\text{a}}(\omega',t)\tau_3\hat{\Gamma}_{\mathrm{N}}\left(\omega,t\right)\hat{G}_0^{\text{r}}(\omega',t)\right.\nonumber\\
&& \left.\frac{\hat{\Sigma}^{\mathrm{corr},<}\left(\omega',t\right)+f\left(\omega'\right)\left(\hat{\Sigma}^{\mathrm{corr},r}\left(\omega',t\right)-\hat{\Sigma}^{\mathrm{corr},a}\left(\omega',t\right)\right)}{\omega'-\omega-i0_+}\right].\nonumber
\end{eqnarray}

It is important to point out that the average-time approximation is exact whenever: i) the dot is non-interacting, ii) temperature is zero, the interaction is arbitrary, and the system can be effectively mapped to a Fermi liquid, and iii) when interaction is infinite and the self-energy $\Sigma$ is calculated up to linear order in $\Gamma$. Therefore the correction term  equals zero in all cases discussed in this paper.

In the following we are interested in calculating the charge pumped through the system  in the weak pumping regime (bilinear response in the pumping fields \cite{note_bil}). The pumped charge $Q$ is related to the pumped current by $Q=\int_{0}^{\mathcal{T}}J\left(t\right)$, where $\mathcal{T}$ is the period of the cycle. We can choose as pumping parameters, $X(t)$ and $Y(t)$, any two of the three quantities $\epsilon\left(t\right)=\bar{\epsilon}+\delta\epsilon\left(t\right)$, $\Gamma_{\mathrm{S}}\left(t\right)=\bar{\Gamma}_{\mathrm{S}}+\delta\Gamma_{\mathrm{S}}\left(t\right)$ and $\Gamma_{\mathrm{N}}\left(t\right)=\bar{\Gamma}_{\mathrm{N}}+\delta\Gamma_{\mathrm{N}}\left(t\right)$. The time-averaged part $\bar{X}$ is denoted by a bar 
and the time-dependent part by $\delta X\left(t\right)$.
For the pumped charge, with pumping parameters indicated in the subscript, we find 
\begin{eqnarray}\label{eqn_weak_pumping}
Q_{X,Y} & & =  -\frac{e\eta\left(X,Y\right)}{4\pi}\int d\omega\left(-\frac{\partial f}{\partial\omega}\right)\mathbb{R}\mathrm{e}\mathrm{Tr}\tau_3\nonumber\\
&&\left\{\frac{\partial}{\partial\bar{ X}}\left(\left(\bar{\Gamma}_{\mathrm{N}}-\bar{\Gamma}_{\mathrm{S}}\right)\bar{G}_0^{\mathrm{r}}\right)\frac{\partial}{\partial \bar{Y}}\left(\left( \bar{G}_0^{\mathrm{r}}\right)^{-1} \bar{G}_0^{\mathrm{a}}\right)\right.\nonumber\\
&&\left.-\frac{\partial}{\partial \bar{Y}}\left(\left(\bar{\Gamma}_{\mathrm{N}}-\bar{\Gamma}_{\mathrm{S}}\right)\bar{G}_0^{\mathrm{r}}\right)\frac{\partial}{\partial \bar{X}}\left(\left( \bar{G}_0^{\mathrm{r}}\right)^{-1} \bar{G}_0^{\mathrm{a}}\right)\right\} .
\end{eqnarray}
The prefactor $\eta\left(X,Y\right)$ accounts for the amplitude and the relative phase of the pumping parameters and is equal to the surface in parameter space enclosed in one pumping cycle. It is defined as $\eta\left(X,Y\right)=\int_0^{\mathcal{T}} d\tau\left[\frac{d}{d\tau}\delta X(\tau)\right]\delta Y(\tau)$. The order of the parameters in the argument of $\eta$ is important and has to be respected in the formulas above (changing their order results in an additional minus sign.) 
$\bar{G}_0^{\mathrm{r}}$ and $\bar{G}_0^{\mathrm{a}}$ are 
the retarded and advanced dot Green's functions (both matrices in Nambu space although the caret has been omitted to simplify the notation) with all parameters  taken at their time-averaged value.

\section{Noninteracting quantum dot}

We start by considering the simple case of a non-interacting quantum dot. The results which we obtain for the non-interacting dot, can equivalently be calculated by a generalization of Brouwer's formula to the N-dot-S system.\cite{wang01,blaauboer02,wang02} The instantaneous Green's function of the dot is 
\begin{equation}\label{eq_G_nonint}
\hat{G}^{\text{r}}(\omega)=\left(\omega-\epsilon\tau_3+\frac{1}{2} i 
\Gamma_{\text{N}}-\Sigma_{\text{S}}\right)^{-1},
\end{equation}
where the self energy due to the proximity of the superconducting lead 
reads
\begin{equation}\label{eq_sigma_nonint}
\hat{\Sigma}_{\text{S}}=\left(\begin{array}{c c}
    -\frac{\Gamma_\mathrm{S}}{2}\frac{\omega}{\sqrt{\Delta^2-\omega^2}} & \frac{\Gamma_\mathrm{S}}{2}\frac{\Delta}{\sqrt{\Delta^2-\omega^2}}\\
   \frac{\Gamma_\mathrm{S}}{2}\frac{\Delta}{\sqrt{\Delta^2-\omega^2}} &   -\frac{\Gamma_\mathrm{S}}{2}\frac{\omega}{\sqrt{\Delta^2-\omega^2}}
\end{array}\right)\ .
\end{equation}
We now focus on the 
limit of large superconducting gap $\Delta \rightarrow \infty$, when 
$(\hat{\Sigma}_\text{S})_{1,2}=(\hat{\Sigma}_\text{S})_{2,1}=\Gamma_{\mathrm{S}}/2$ and $(\hat{\Sigma}_\text{S})_{1,1}=(\hat{\Sigma}_\text{S})_{2,2}=0$.

Let us consider first the particle-particle spectral function, which is defined as 
$A_{1,1}(\omega)=-(1/\pi)\mathbb{I}\mathrm{m}\left\{G^{\text{r}}_{1,1}\right\}$. 
The spectral function conveys important information on the proximity effect in the quantum dot. Furthermore, the DC transport properties are influenced by the energy distribution of spectral weight, for example the  Andreev-reflection probability $R_{\text{A}}(\omega)$ has a similar structure as   $A_{1,1}(\omega)$. Due to the close relation of the spectral function with the retarded Green's function appearing in the formula for the pumped charge, the spectral function will also prove to be a useful quantity to analyze pumping in the N-dot-S system. 
It turns out that $A_{1,1}$ is given by a linear combination of two Lorentzians and reads
\begin{eqnarray}
\nonumber
 A_{1,1}(\omega)  = & \frac{1}{2}\left(1+\frac{\epsilon}{\sqrt{\frac{\Gamma_{\mathrm{S}}^2}{4}+\epsilon^2}}\right)
L_{\Gamma_{\mathrm{N}}}\left(\omega-\sqrt{\frac{\Gamma_{\mathrm{S}}^2}{4}+\epsilon^2} \right)\\
 + & 
\frac{1}{2}\left(1-\frac{\epsilon}{\sqrt{\frac{\Gamma_{\mathrm{S}}^2}{4}+\epsilon^2}}\right)
L_{\Gamma_{\mathrm{N}}}\left(\omega+\sqrt{\frac{\Gamma_{\mathrm{S}}^2}{4}+\epsilon^2} \right),
\label{eq_spectral_nonint}
\end{eqnarray}
with the Lorentzian bell defined by 
$L_{\Gamma}(\omega-E)= (\frac{\Gamma}{2\pi})/\left(\left(\omega-E\right)^2+\left(\frac{\Gamma}{2}\right)^2\right)$. 
The presence of two peaks in the particle-particle spectral function [Eq.~(\ref{eq_spectral_nonint})] is a signature of the proximity effect induced by the superconductor. 
The width of the Lorentzians appearing in Eq.~(\ref{eq_spectral_nonint}) is equal to the coupling to the normal lead, while their positions are shifted, with respect to the dot-level position, by the coupling to the superconducting lead $\Gamma_{\mathrm{S}}$.
This fact can be referred to as a {\em proximization} of the dot-level position, which might be contrasted to the level {\em renormalization} due to Coulomb interaction. 
The repulsion of the two Lorentzian peaks with increasing $\Gamma_{\mathrm{S}}$ reflects the increasing coupling between electron and hole excitations in the dot [represented by the off-diagonal terms in Eq. (\ref{eq_sigma_nonint})].

In the non-interacting case Eq. (\ref{eqn_weak_pumping}) can be written in a compact form if the two pumping parameters, denoted by $X\left(t\right)$ and $Y\left(t\right)$, are chosen among the three quantities $\left\{\frac{\Gamma_{\mathrm{N}}}{2}\left(t\right),\frac{\Gamma_{\mathrm{S}}}{2}\left(t\right),\epsilon\left(t\right)\right\}$.\cite{note2}
We denote by $Z$ the parameter which is kept constant in time. Eq. (\ref{eqn_weak_pumping}) simplifies to 
\begin{eqnarray}\label{eqn_charge_nonint}
Q_{X,Y} & = & \frac{4e}{\pi}\eta\left(X,Y\right)\varepsilon_{X,Y,Z}\\
&&\int d\omega\left(-\frac{\partial f}{\partial\omega}\right)\frac{\bar{\Gamma}_{\mathrm{S}}}{2}\left(\frac{\bar{\Gamma}_{\mathrm{N}}}{2}\right)^2\frac{\partial}{\partial \bar{Z}}\left(\frac{1}{N}\right)\nonumber\ ,
\end{eqnarray}
where $N$ is the denominator of the particle-particle spectral function with time-averaged parameters
\begin{eqnarray}
N & = & \left\{\omega^2-\bar{\epsilon}^2-\left[\left(
\frac{\bar{\Gamma}_{\mathrm{S}}}{2}\right)^2-\left(
\frac{\bar{\Gamma}_{\mathrm{N}}}{2}\right)^2\right]\right\}^2\\
& + & 4\left(\frac{\bar{\Gamma}_{\mathrm{N}}}{2}\right)^2\left[\left(\frac{\bar{\Gamma}_{\mathrm{S}}}{2}\right)^2+\bar{\epsilon}^2\right] \ .\nonumber
\end{eqnarray} 
The quantity $\eta\left(X,Y\right)$ is the surface in parameter space enclosed in one pumping cycle. The antisymmetric tensor $\varepsilon_{X,Y,Z}$ has to be evaluated respecting the order of the parameters $\left\{\frac{\Gamma_{\mathrm{N}}}{2}\left(t\right),\frac{\Gamma_{\mathrm{S}}}{2}\left(t\right),\epsilon\left(t\right)\right\}$. 
In the case of weak-pumping with the two tunnel-barrier strengths, i.e $X(t),Y(t)=
\bar{\Gamma}_{\text{S,N}}+\delta\Gamma_{\text{S,N}}(t)$, and $\bar{Z}=\bar{\epsilon}$, we can rewrite Eq. (\ref{eqn_charge_nonint}) as 
\begin{equation}
\label{q_delta_0}
Q_{\Gamma_{\mathrm{N}},\Gamma_{\mathrm{S}}}=\frac{e}{ \pi}\eta\left(\Gamma_{\mathrm{N}},\Gamma_{\mathrm{S}}\right) \frac{1} {2 \bar{\Gamma}_{\mathrm{S}}}  \int d\omega \left(- \frac{\partial f}
  {\partial \omega} \right) \frac{\partial \bar{R}_{\text{A}}}{\partial \bar{\epsilon}}, 
\end{equation} 
which relates the pumped charge to the probability of Andreev reflection $\bar{R}_{\text{A}}$,
computed with the time-dependent parameters frozen at their average value. 
The probability of Andreev reflection is given by\cite{beenakker}
\begin{eqnarray}
\nonumber
R_{\text{A}}&=&\frac{\Gamma^2_{\mathrm{N}}\Gamma^2_{\mathrm{S}}}
{4\left[\omega^2-\epsilon^2+\frac{1}{4}\left(\Gamma^2_{\mathrm{N}}-\Gamma^2_{\mathrm{S}}
\right)\right]^2+\Gamma^2_{\mathrm{N}}\left(\Gamma^2_{\mathrm{S}} +4\epsilon^2\right)}\\
&=&\frac{\pi}{4}\frac{\Gamma_{\mathrm{N}}\Gamma_{\mathrm{S}}^2}{\frac{\Gamma_{\mathrm{S}}^2}{4}+\frac{\Gamma_{\mathrm{N}}^2}{4}+\epsilon^2+\omega^2}\left[
L_{\Gamma_{\mathrm{N}}}\left(\omega-\sqrt{\frac{\Gamma_{\mathrm{S}}^2}{4}+\epsilon^2} \right)\right.\nonumber\\
& + & \left. L_{\Gamma_{\mathrm{N}}}\left(\omega+\sqrt{\frac{\Gamma_{\mathrm{S}}^2}{4}+\epsilon^2} \right)
\right].
\label{andref}
\end{eqnarray}\\

\begin{figure}
\includegraphics[width=2.5in]{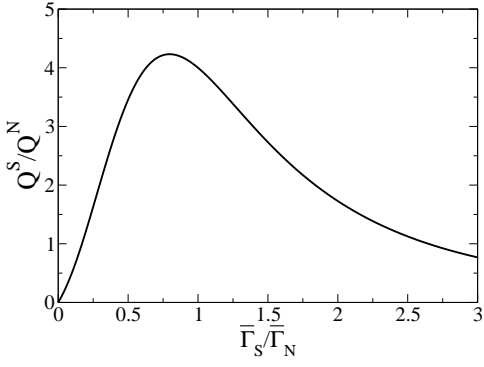}
\caption{Ratio between the pumped charge in the N-dot-S and in the N-dot-N systems, $Q^{\mathrm{S}}_{X,Y}/Q^{\mathrm{N}}_{X,Y}$,  as a function of $\bar{\Gamma}_{\mathrm{S}}/\bar{\Gamma}_{\mathrm{N}}$. The curve is the same for any couple of pumping parameters $X,Y$. 
Temperature is equal to zero and the bare dot level is at resonance.
\label{fig_ratio_nonint}}
\end{figure}

\begin{figure}
\includegraphics[width=2.5in]{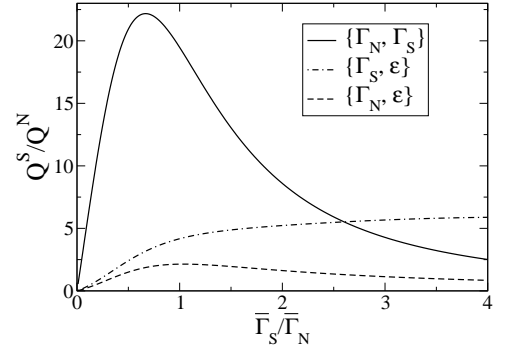}
\caption{Ratio between the pumped charge in N-dot-S and in the N-dot-N system, 
$Q^{\mathrm{S}}_{X,Y}/Q^{\mathrm{N}}_{X,Y}$, as a function of $\bar{\Gamma}_{\mathrm{S}}/\bar{\Gamma}_{\mathrm{N}}$. Temperature is equal to $k_{\mathrm{B}}T=1.5\bar{\Gamma}_{\mathrm{N}}$. The pumping parameters are $\Gamma_{\mathrm{N}}$ and  $\Gamma_{\mathrm{S}}$ (solid line),  $\Gamma_{\mathrm{N}}$ and $\epsilon$ (dashed line) and $\Gamma_{\mathrm{S}}$ and $\epsilon$ (dotted and dashed line). The bare   dot level is at resonance.
\label{fig_ratio_nonint_Tfinite}}
\end{figure}

We are now interested in comparing our results to the ones obtained for an N-dot-N system (see Fig. \ref{fig_system} with normal-conducting right lead).
The current pumped through a non-interacting dot attached to two normal conducting leads 
can be obtained using Brouwer's formula.\cite{brouwer98} 
For the case where the tunneling rates $\Gamma_{\mathrm{N}}(t)$ and  $\Gamma_{\mathrm{S}}(t)$  are the pumping parameters, it has been found that the current in the presence of one superconducting lead is about a factor 4 bigger than in the case of two normal leads, if the dot level is at resonance ($\epsilon=0$) and temperature is zero.\cite{blaauboer02, wang01,wang02}
We find exactly the same behavior for all three choices of pumping parameters, at zero temperature.
It turns out that the ratio $Q^{\mathrm{S}}_{X,Y}/Q^{\mathrm{N}}_{X,Y}$ (equal for any pairs of pumping parameters) has a maximum, as a function of the level position, at resonance.\cite{note1} 
We plot in Fig.~\ref{fig_ratio_nonint} the ratio $Q^{\mathrm{S}}_{X,Y}/Q^{\mathrm{N}}_{X,Y}$, for $\epsilon=0$, as a function of $\bar{\Gamma}_{\mathrm{S}}/\bar{\Gamma}_{\mathrm{N}}$: a maximum is present when $\bar{\Gamma}_{\mathrm{S}}/\bar{\Gamma}_{\mathrm{N}}\simeq 0.9$.
In the limit of small and large $\bar{\Gamma}_{\mathrm{S}}/\bar{\Gamma}_{\mathrm{N}}$ the charge ratio goes to zero.
This happens because, in the presence of a superconducting lead, pumping is mediated by the Andreev reflection probability, Eq.~(\ref{andref}).
On one hand, for $\bar{\Gamma}_{\mathrm{S}}/\bar{\Gamma}_{\mathrm{N}}\ll 1$, the ratio is suppressed since $R_{\mathrm{A}}$ implies higher-order tunneling processes to the superconducting lead with respect to tunneling to the normal lead.
On the other hand, for $\bar{\Gamma}_{\mathrm{S}}/\bar{\Gamma}_{\mathrm{N}}\gg1$, the Lorentzian contributions to $R_{\mathrm{A}}$ get narrower with decreasing $\bar{\Gamma}_{\mathrm{N}}$ and more distant with increasing $\bar{\Gamma}_{\mathrm{S}}$, thus suppressing the Andreev reflection probability at $\omega=0$ (relevant for zero temperature). 
In the case of finite temperature the charge ratio depends on the parameter choice and can take values much bigger than $4$. Fig. \ref{fig_ratio_nonint_Tfinite} shows the ratio $Q^{\mathrm{S}}_{X,Y}/Q^{\mathrm{N}}_{X,Y}$ for the three different parameter choices at temperature $k_{\mathrm{B}}T=1.5\bar{\Gamma}_{\mathrm{N}}$, with the dot level being at resonance.

In the following we discuss some qualitative features, in the high-temperature limit ($k_{\text{B}} T \gg \Gamma_{\text{N}}$), of the ratio $Q^{\mathrm{S}}_{X,Y}/Q^{\mathrm{N}}_{X,Y}$. Expressions for the pumped charge are reported in Appendix \ref{A1}.  
While for the N-dot-S system the leading order of
the pumped charge is the first derivative of the Fermi function, in the N-dot-N system this is the case only when the level position is one of the pumping parameters. The first contribution to $Q^{\mathrm{N}}_{\Gamma_\mathrm{N},\Gamma_\mathrm{S}}$ starts with the second derivative of the Fermi function.
On the other hand, the parameter $\Gamma_\mathrm{S}$ enters in the spectral density of the N-dot-S system via the proximized peak positions.
The high-temperature limit of the charge ratio as a function of temperature is therefore a constant  if $\{\Gamma_{\alpha},\epsilon\}$ are the pumping parameters and it is quadratic in temperature in the case that $\{\Gamma_\mathrm{N},\Gamma_\mathrm{S}\}$ are the pumping parameters:
\begin{subequations}
\label{eq_ratio_highT}
\begin{eqnarray}
\frac{Q^{\mathrm{S}}_{\Gamma_\mathrm{N},\Gamma_\mathrm{S}}}{Q^{\mathrm{N}}_{\Gamma_\mathrm{N},\Gamma_\mathrm{S}}} & \simeq &\left(k_{\mathrm{B}}T\right)^2\frac{\bar{\Gamma}_\mathrm{N}\bar{\Gamma}_\mathrm{S}}{\left(\bar{\epsilon}^2+\frac{\bar{\Gamma}_\mathrm{N}^2}{4}+\frac{\bar{\Gamma}_\mathrm{S}^2}{4}\right)^2}\label{eq_ratio_highT_1} 
\end{eqnarray}
\begin{eqnarray}
\frac{Q^{\mathrm{S}}_{\Gamma_\mathrm{N},\epsilon}}{Q^{\mathrm{N}}_{\Gamma_\mathrm{N},\epsilon}} & \simeq & \frac{1}{8}\frac{\bar{\Gamma}_\mathrm{N}\bar{\Gamma}_\mathrm{S}\left(\bar{\Gamma}_\mathrm{N}+\bar{\Gamma}_\mathrm{S}\right)^2}{\left(\bar{\epsilon}^2+\frac{\bar{\Gamma}_\mathrm{N}^2}{4}+\frac{\bar{\Gamma}_\mathrm{S}^2}{4}\right)^2}\label{eq_ratio_highT_2}
\end{eqnarray}
\begin{eqnarray}
\frac{Q^{\mathrm{S}}_{\Gamma_\mathrm{S},\epsilon}}{Q^{\mathrm{N}}_{\Gamma_\mathrm{S},\epsilon}} & \simeq & \frac{1}{8}\frac{\bar{\Gamma}_\mathrm{N}\bar{\Gamma}_\mathrm{S}\left(\bar{\Gamma}_\mathrm{N}+\bar{\Gamma}_\mathrm{S}\right)^2}{\bar{\epsilon}^2+\frac{\bar{\Gamma}_\mathrm{N}^2}{4}+\frac{\bar{\Gamma}_\mathrm{S}^2}{4}}\left(\frac{1}{\bar{\epsilon}^2+\frac{\bar{\Gamma}_\mathrm{N}^2}{4}+\frac{\bar{\Gamma}_\mathrm{S}^2}{4}}+\frac{2}{\bar{\Gamma}_\mathrm{N}^2}\right) .
\nonumber\\
\label{eq_ratio_highT_3}
\end{eqnarray}
\end{subequations}
For all three cases described by Eqs.~(\ref{eq_ratio_highT}), the ratio of the pumped charge, as a function of the average dot level position, has a maximum when the average dot level is at resonance.
At $\bar{\epsilon}=0$, with $\bar{\Gamma}_\mathrm{S}=\bar{\Gamma}_\mathrm{N}$,
we find that $Q^{\mathrm{S}}_{\Gamma_\mathrm{N},\epsilon}/Q^{\mathrm{N}}_{\Gamma_\mathrm{N},\epsilon} \simeq 2$ and $Q^{\mathrm{S}}_{\Gamma_\mathrm{S},\epsilon}/Q^{\mathrm{N}}_{\Gamma_\mathrm{S},\epsilon} \simeq 4$. 
Furthermore we note that due to the second term in Eq. (\ref{eq_ratio_highT_3}), the ratio $Q^{\mathrm{S}}_{\Gamma_\mathrm{S},\epsilon}/Q^{\mathrm{N}}_{\Gamma_\mathrm{S},\epsilon}$ is increasing in a monotonic way for increasing $\bar{\Gamma}_\mathrm{S}/\bar{\Gamma}_\mathrm{N}$, while $Q^{\mathrm{S}}_{\Gamma_\mathrm{N},\Gamma_\mathrm{S}}/Q^{\mathrm{N}}_{\Gamma_\mathrm{N},\Gamma_\mathrm{S}}$ and $Q^{\mathrm{S}}_{\Gamma_\mathrm{N},\epsilon}/Q^{\mathrm{N}}_{\Gamma_\mathrm{N},\epsilon}$ have a maximum as a function of $\bar{\Gamma}_\mathrm{S}/\bar{\Gamma}_\mathrm{N}$.
This behavior is present in Fig. \ref{fig_ratio_nonint_Tfinite}, where we also observe a large value of the maximum of $Q^{\mathrm{S}}_{\Gamma_\mathrm{N},\Gamma_\mathrm{S}}/Q^{\mathrm{N}}_{\Gamma_\mathrm{N},\Gamma_\mathrm{S}}$, which is due to the quadratic dependence on $k_{\mathrm{B}}T$.

\begin{figure}
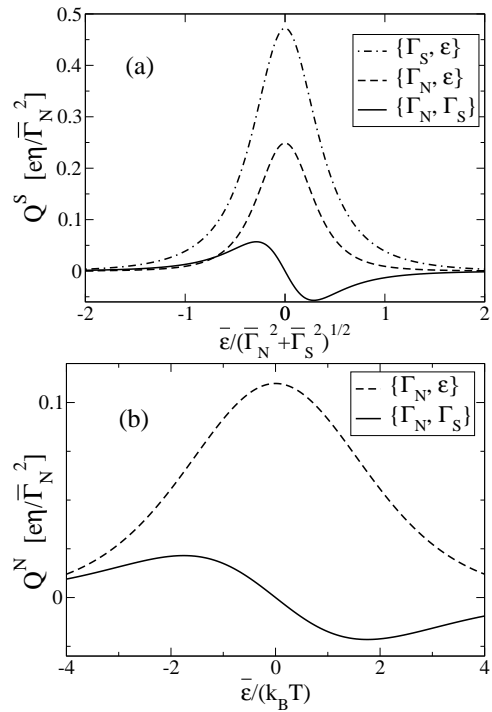

\includegraphics[width=2.5in]{fig4}
\includegraphics[width=2.5in]{fig5}
\caption{Pumped charge in the N-dot-S [panel (a)] and in the N-dot-N [panel (b)] systems as a function of the average level position $\bar{\epsilon}$. Temperature is set to $k_{\mathrm{B}}T=\bar{\Gamma}_{\mathrm{N}}$ and $\bar{\Gamma}_{\mathrm{N}}=\bar{\Gamma}_{\mathrm{S}}$. The pumping parameters are $\{\Gamma_{\mathrm{N}},\,\Gamma_{\mathrm{S}}\}$ (solid line),  $\{\Gamma_{\mathrm{N}},\,\epsilon\}$ (dashed line) and $\{\Gamma_{\mathrm{S}},\,\epsilon\}$ (dashed-dotted line). The units on the x-axes are the ones determining the width of the respective function. 
\label{fig_charge_nonint}}
\end{figure}

The plots in Fig. \ref{fig_charge_nonint} show the pumped charge, as a function of the average dot level position, through the N-dot-S system [panel (a)] and through the N-dot-N system [panel (b)] for different choices of pumping-parameter pairs at $k_{\mathrm{B}}T=\bar{\Gamma}_{\mathrm{N}}$ and for $\bar{\Gamma}_{\mathrm{N}}=\bar{\Gamma}_{\mathrm{S}}$.
We point out several differences: In the N-dot-N case the pumped charge is the same for pumping with $\{\Gamma_{\mathrm{N}},\epsilon\}$ and with $\{\Gamma_{\mathrm{S}},\epsilon\}$.
In the presence of a superconducting lead the symmetry between the two leads is absent as soon as temperature differs from zero. Furthermore we report that the width of the pumped charge is strongly temperature dependent for the N-dot-N system, while for the N-dot-S system the width is saturating for high temperatures to a value that depends solely on the coupling to the leads, $\Gamma_{\mathrm{N}}$ and $\Gamma_{\mathrm{S}}$. This is due to the fact that the gap of the superconductor is always bigger than temperature and no quasiparticle transport, which is sensitive to temperature, takes place between the superconductor and the dot.
The symmetry of the pumped charge as a function of the dot-level position is present for both systems.
For the N-dot-S system, when pumping with $\Gamma_{\mathrm{N}}$ and $\Gamma_{\mathrm{S}}$, this happens despite the fact that, at least at high temperatures, pumping is dominated by the proximized level position.

It might be interesting to contrast the pumped charge in the N-dot-S system with the DC linear conductance of the same system (here all parameters are time-independent):
\begin{equation}
G=\frac{4e^2}{h} \int d\omega \left(- \frac{\partial f}
  {\partial \omega} \right) R_{\text{A}}(\omega). 
\end{equation}
In the high temperature limit, i.e. $ k_{\text{B}} T\gg \Gamma_{\mathrm{N}}$, the 
conductance is given by 
\begin{equation}
\label{dc-noninteracting}
G=-\frac{4e^2}{h} \frac{\pi}{4}
\frac{\Gamma_{\mathrm{N}}\Gamma^2_{\mathrm{S}}}
{
\frac{\Gamma_{\mathrm{S}}^2}{4}+\frac{\Gamma_{\mathrm{N}}^2}{4}+\epsilon^2}  \frac{\partial f}
  {\partial \omega}\left(\sqrt{\frac{\Gamma_{\mathrm{S}}^2}{4}+\epsilon^2} \right).
\end{equation}

\section{Strong interaction limit}

In the regime of strong Coulomb repulsion double occupation is forbidden 
and coherent Andreev scattering can occur only if $\Delta$ is finite (electrons can enter the superconductor within a time window of order 
$\hbar/\Delta$).
In the Kondo limit, when temperature is zero, we can evaluate the retarded equilibrium dot Green's function making use of the mean-field slave-boson technique.\cite{coleman84,schwab99} 
As the dot can only be empty or singly occupied we can describe the singly-occupied dot state by the fermion operator $f_{\sigma}$ (known as pseudo fermion) and the empty dot by the boson operator $b$. The real dot operator $d_{\sigma}$ is simply given by $d_{\sigma}=b^{\dagger}f_{\sigma}$. 
The operators $b$ and $f_{\sigma}$ need to fulfill the constraint $b^{\dagger}b+\sum_{\sigma}f_{\sigma}^{\dagger}f^{}_{\sigma}=1$. Such a constraint is introduced in the Hamiltonian using the Lagrange multiplier $\lambda$ which plays the role of a chemical potential.  
In the mean-field limit, the  boson operator $b$ is replaced by a real number $b_{0}$.
The quantities $b_0$ and $\lambda$ are determined by minimizing the free energy. 
Note that we are allowed to use an equilibrium formalism, since after performing the adiabatic approximation we are left with a current formula, Eq. (\ref{adiabaticcurrent}), that contains only equilibrium quantities. 

Details of the derivation of the retarded dot Green's function by means of the slave-boson 
technique, in the limit $\Delta\gg \Gamma_{\text{S}}$, are given in Appendix \ref{A2}.
It is important to notice that the pumped charge in the deep Kondo regime is zero.
Therefore, in the following, we will be interested in the mixed-valence regime only.  
\begin{figure}
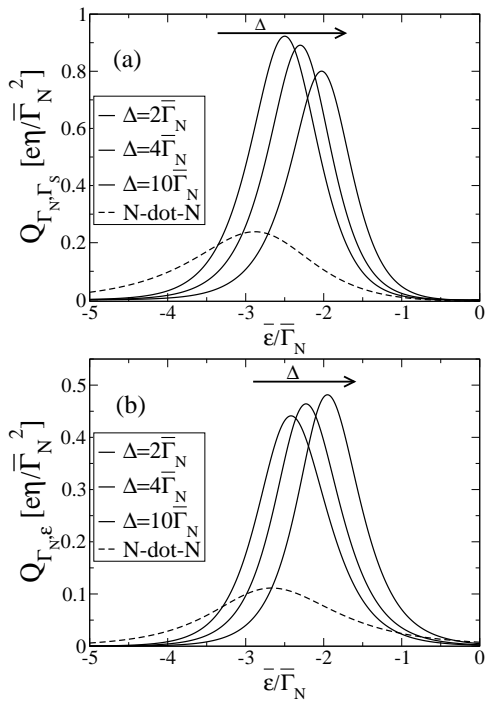

\includegraphics[width=2.5in]{fig6}
\includegraphics[width=2.5in]{fig7}
\caption{
Pumped charge  $Q^{\mathrm{S}}_{\Gamma_{\mathrm{N}},\Gamma_{\mathrm{S}}}$ [panel (a)] and $Q^{\mathrm{S}}_{\epsilon,\Gamma_{\mathrm{S}}}$ [panel (b)] as a function of the average level position for different values of $\Delta$
, solid lines, and $Q^{\mathrm{N}}_{\Gamma_{\mathrm{N}},\Gamma_{\mathrm{S}}}$  and $Q^{\mathrm{N}}_{\Gamma_{\mathrm{N}},\epsilon}$, dashed lines. Temperature is zero and we consider symmetric barriers ($\bar{\Gamma}_{\text{N}}=\bar{\Gamma}_{\text{S}}$). The cut-off energy is $W=20\bar{\Gamma}_{\mathrm{N}}$.
\label{fig_delta_dip}}
\end{figure}
The pumped charge in the weak-pumping limit for the aforementioned three pairs of pumping parameters can be calculated using Eq. (\ref{eqn_weak_pumping}). The pumped charge has a similar structure in all three cases.
We find, in particular, the general relation 
\begin{equation}\label{eq_similar_charge}
-\frac{1}{\eta\left(\Gamma_{\mathrm{N}},\epsilon\right)}\frac{1}{\bar{\Gamma}_{\mathrm{S}}}Q^{\mathrm{S}}_{\Gamma_{\mathrm{N}},\epsilon}=\frac{1}{\eta\left(\Gamma_{\mathrm{S}},\epsilon\right)}\frac{1}{\bar{\Gamma}_{\mathrm{N}}}Q^{\mathrm{S}}_{\Gamma_{\mathrm{S}},\epsilon} \ ,
\end{equation}
which is also valid for the noninteracting N-dot-S system at zero temperature.

In analyzing the properties of the pumped charge as a function of the different system parameters 
we can profit from the fact that, in the Kondo regime, the system behaves like a non-interacting system with renormalized parameters, denoted as $\tilde{\Gamma}_\mathrm{N}$, $\tilde{\Gamma}_\mathrm{S}$ and $\tilde{\epsilon}$. We should however keep in mind that the renormalized 
parameters depend in a complex way on all the bare parameters.
Panel (a) of Fig. \ref{fig_delta_dip} shows the pumped charge $Q^{\mathrm{S}}_{\Gamma_{\mathrm{N}},\Gamma_{\mathrm{S}}}$ (solid lines) 
 as a function of the level position, for increasing values of the superconducting gap $\Delta$. The charge  $Q^{\mathrm{N}}_{\Gamma_{\mathrm{N}},\Gamma_{\mathrm{S}}}$ in the N-dot-N is also shown for reference (dashed line). In panel (b), the pumped charge is shown for 
$\Gamma_{\mathrm{N}}$ and $\epsilon$ chosen as pumping parameters. We note that for both the N-dot-N system and the N-dot-S system (for any value of the superconducting gap $\Delta$) the charge has a bell-like structure.
The width of the bell-like curves reflects the width of the respective particle-particle spectral functions, which for the N-dot-S system is given by $\tilde{\Gamma}_\mathrm{N}$ while for the
N-dot-N system it is given by $\tilde{\Gamma}_\mathrm{N}+\tilde{\Gamma}_\mathrm{S}$. 
For the values of parameters considered in Fig. \ref{fig_delta_dip} ($\bar{\Gamma}_{\mathrm{N}}=\bar{\Gamma}_{\mathrm{S}}=0.05 W$), the curve relative to $\Delta=2\Gamma_{\mathrm{N}}$ is still very far away from the curve relative to the N-dot-N system.

\begin{figure}
\includegraphics[width=2.5in]{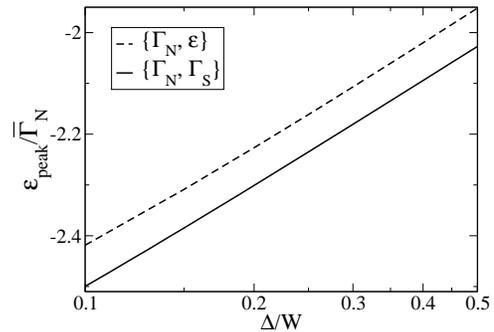}\\
\caption{
The peak position $\varepsilon_{\mathrm{peak}}$ is shown as a function of the superconducting gap for different pairs of pumping parameters on a logarithmic scale. The pumping parameters are $\{\Gamma_{\mathrm{N}},\Gamma_{\mathrm{S}}\}$ (solid line) and $\{\Gamma_{\mathrm{N}},\epsilon\}$ (dashed line). The temperature is zero and $\bar{\Gamma}_{\mathrm{N}}=\bar{\Gamma}_{\mathrm{S}}=0.05 W$.
\label{fig_peakposition}}
\end{figure}

The striking similarity of the pumped charge for the two different pairs of pumping parameters [compare Figs. \ref{fig_delta_dip} (a) and (b) and Eq. (\ref{eq_similar_charge})] can be explained by the fact that pumping takes place via renormalized parameters.
Indeed, in the zero-temperature regime with strong Coulomb interaction, the Green's function of the system depends on the bare parameters solely via renormalized parameters $\tilde{\Gamma}_{\mathrm{N}},\tilde{\Gamma}_{\mathrm{S}}$ and $\tilde{\epsilon}$. 
Therefore, the detailed behavior of the pumped charge is determined by the derivatives of the renormalized parameters with respect to the bare ones.
In particular when the bare dot level is deep below the Fermi energy, i.e. in the Kondo regime, the renormalized parameters are independent of the bare parameters and therefore the pumped charge is zero.

Now, we focus our attention on the dependence of the pumped charge on the superconducting 
gap. In a non-interacting dot, the pumped charge is always independent of the superconducting gap $\Delta$ at $T=0$, since the Green's function of Eq. (\ref{eq_G_nonint}), for $\omega=0$, is independent of $\Delta$. The situation is different in the presence of Coulomb interaction, where   
the pumped charge depends on the superconducting gap via the renormalized parameters.  
Now, we investigate how the features of the bell curves, describing the  pumped charge as a function of the average level $\bar{\epsilon}$, depend on the gap $\Delta$.\\
\textit{Peak position.} Fig. \ref{fig_peakposition} shows that the position of the peak of the pumped charge $Q^{\mathrm{S}}$ depends on $\ln \Delta$. The peak position is shifted towards zero (from below) for increasing gap. This is consistent with the fact that the renormalized level acquires a logarithmic correction in $\Delta$ proportional to $\Gamma_{\mathrm{S}}$.\cite{schwab99} 
\begin{figure}
\includegraphics[width=2.5in]{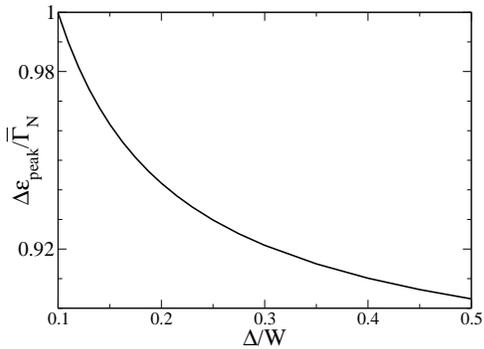}\\
\caption{
Peak width $\Delta\varepsilon_{\mathrm{peak}}$ for the pumped charge plotted as function of the superconducting gap $\Delta$. The pumping parameters are $\{\Gamma_{\mathrm{N}},\Gamma_{\mathrm{S}}\}$. The temperature is zero and $\bar{\Gamma}_{\mathrm{N}}=\bar{\Gamma}_{\mathrm{S}}=0.05 W$.
\label{fig_width}}
\end{figure}

\textit{Peak width.} Fig. \ref{fig_width} shows the peak width $\Delta\varepsilon_{\mathrm{peak}}$ (half-height-width) for the pumped charge $Q^{\mathrm{S}}_{\Gamma_{\mathrm{N}},\Gamma_{\mathrm{S}}}$  as a  function of the superconducting gap $\Delta$. 
As stated before, the peak width has an analogous behavior to the width of the spectral function. 
The coupling to the superconducting lead influences the width of the spectral function via the 
renormalized coupling $\tilde{\Gamma}_{\mathrm{N}}$. The decrease of $\tilde{\Gamma}_{\mathrm{N}}$ with increasing superconducting gap induces the $\Delta$-dependence shown in  Fig.~\ref{fig_width}.

\section{Conclusions}
We have extended a Green's function formalism to compute the pumped current through a quantum dot with Coulomb interaction, to the case when the dot is tunnel-coupled to one normal and one superconducting lead.
First, we have applied our formalism to a non-interacting dot in an N-dot-S configuration, where we derived a compact formula for the pumped charge, and we  calculated the ratio between the charge pumped through the N-dot-S system and through an N-dot-N system. At finite temperature, we found different characteristic behaviors of this ratio, depending on the choice of pumping parameters.
Then, we have studied the pumped charge through a quantum dot with infinitely strong Coulomb interaction at zero temperature using the mean-field slave-boson approach. In particular, 
we have focused our attention on the influence of the superconducting gap $\Delta$ on the 
pumped charge. We have explained features of the pumped charge exploiting the
mapping to a non-interacting system with renormalized parameters.
In the end we wish to comment on the experimental realization of the pumping
mechanism discussed theoretically in this article. 
The system depicted in Fig.~\ref{fig_system} can be realized in analogy to an experiment reported recently by van Dam \textit{et al.}~\cite{vandam06}.
Here a semiconductor quantum dot is contacted with superconductors, using indium arsenide nanowires combined with local gate voltages to create a quantum dot with tunable coupling to the
leads. Superconductivity is induced in the wires by means of the proximity effect using aluminum contacts.
In order to realize pumping, time-dependent control of the system parameters
is required.

\begin{appendix}
\section{Pumped charge in the Non-interacting dot: $k_{\text{B}} T\gg \Gamma_{\text{N}}$} \label{A1}
In this appendix we report the expressions for the pumped charge through the non-interacting 
dot in the weak-pumping regime in the limit $k_{\text{B}} T\gg \Gamma_{\text{N}}$.\\
{\it N-dot-S system} From Eq.~(\ref{eqn_charge_nonint}), for the three
possible choices of pumping fields, we obtain:  
\begin{widetext}
\begin{eqnarray}
Q^{\mathrm{S}}_{\Gamma_\mathrm{N},\Gamma_\mathrm{S}} 
\simeq
 e\eta\left(\Gamma_\mathrm{N},\Gamma_\mathrm{S}\right)\frac{\bar{\epsilon}\bar{\Gamma}_\mathrm{N}\bar{\Gamma}_\mathrm{S}}{\bar{\epsilon}^2+\frac{\bar{\Gamma}_\mathrm{N}^2}{4}+\frac{\bar{\Gamma}_\mathrm{S}^2}{4}}
\left[\frac{1}{4}\frac{1}{\bar{\epsilon}^2+\frac{\bar{\Gamma}_\mathrm{N}^2}{4}+\frac{\bar{\Gamma}_\mathrm{S}^2}{4}}\frac{\partial
    f}{\partial
    \omega}\left(\sqrt{\bar{\epsilon}^2+\frac{\bar{\Gamma}_\mathrm{S}^2}{4}}\right)
-\frac{1}{8}\frac{1}{\sqrt{\bar{\epsilon}^2+\frac{\bar{\Gamma}_\mathrm{S}^2}{4}}}\frac{\partial^2 f}{\partial \omega^2}\left(\sqrt{\bar{\epsilon}^2+\frac{\bar{\Gamma}_\mathrm{S}^2}{4}}\right)\right] 
\label{eq_highT_1}
\end{eqnarray}
\begin{eqnarray}
Q^{\mathrm{S}}_{\Gamma_\mathrm{N},\epsilon}
\simeq
 -e\eta\left(\Gamma_\mathrm{N},\epsilon\right)\frac{\bar{\Gamma}_\mathrm{N}\bar{\Gamma}_\mathrm{S}^2}{\bar{\epsilon}^2+\frac{\bar{\Gamma}_\mathrm{N}^2}{4}+\frac{\bar{\Gamma}_\mathrm{S}^2}{4}}
\left[\frac{1}{4}\frac{1}{\bar{\epsilon}^2+\frac{\bar{\Gamma}_\mathrm{N}^2}{4}+\frac{\bar{\Gamma}_\mathrm{S}^2}{4}}\frac{\partial
    f}{\partial
    \omega}\left(\sqrt{\bar{\epsilon}^2+\frac{\bar{\Gamma}_\mathrm{S}^2}{4}}\right)
-\frac{1}{8}\frac{1}{\sqrt{\bar{\epsilon}^2+\frac{\bar{\Gamma}_\mathrm{S}^2}{4}}}\frac{\partial^2 f}{\partial \omega^2}\left(\sqrt{\bar{\epsilon}^2+\frac{\bar{\Gamma}_\mathrm{S}^2}{4}}\right)\right]\label{eq_highT_2}
\end{eqnarray}
\begin{eqnarray}
Q^{\mathrm{S}}_{\Gamma_\mathrm{S},\epsilon} 
 \simeq 
 e\eta\left(\Gamma_\mathrm{S},\epsilon\right)\frac{\bar{\Gamma}_\mathrm{N}^2\bar{\Gamma}_\mathrm{S}}{\bar{\epsilon}^2+\frac{\bar{\Gamma}_\mathrm{N}^2}{4}+\frac{\bar{\Gamma}_\mathrm{S}^2}{4}}
\left[\left(\frac{1}{4}\frac{1}{\bar{\epsilon}^2+\frac{\bar{\Gamma}_\mathrm{N}^2}{4}+\frac{\bar{\Gamma}_\mathrm{S}^2}{4}}+\frac{1}{2\bar{\Gamma}_\mathrm{N}^2}\right)\frac{\partial
    f}{\partial
    \omega}\left(\sqrt{\bar{\epsilon}^2+\frac{\bar{\Gamma}_\mathrm{S}^2}{4}}\right)
-\frac{1}{8}\frac{1}{\sqrt{\bar{\epsilon}^2+\frac{\bar{\Gamma}_\mathrm{S}^2}{4}}}\frac{\partial^2 f}{\partial \omega^2}\left(\sqrt{\bar{\epsilon}^2+\frac{\bar{\Gamma}_\mathrm{S}^2}{4}}\right)\right] .
\label{eq_highT_3}
\end{eqnarray}
\end{widetext}
{\it N-dot-N system} We compare the results for the N-dot-S system to the charge pumped through the N-dot-N system by varying the respective pumping parameters. We find:
\begin{eqnarray}
Q^{\mathrm{N}}_{\Gamma_\mathrm{N},\Gamma_\mathrm{S}} & \simeq & -\frac{e}{2}\eta\left(\Gamma_\mathrm{N},\Gamma_\mathrm{S}\right)\frac{\partial^2f}{\partial\omega^2}\left(\bar{\epsilon}\right)\\
Q^{\mathrm{N}}_{\Gamma_\mathrm{N},\epsilon} & \simeq & -2e\eta\left(\Gamma_\mathrm{N},\epsilon\right)\frac{\bar{\Gamma}_\mathrm{S}}{\left(\bar{\Gamma}_\mathrm{N}+\bar{\Gamma}_\mathrm{S}\right)^2}\frac{\partial f}{\partial\omega}\left(\bar{\epsilon}\right)\\
Q^{\mathrm{N}}_{\Gamma_\mathrm{S},\epsilon} & \simeq & 2e\eta\left(\Gamma_\mathrm{S},\epsilon\right)\frac{\bar{\Gamma}_\mathrm{N}}{\left(\bar{\Gamma}_\mathrm{N}+\bar{\Gamma}_\mathrm{S}\right)^2}\frac{\partial f}{\partial\omega}\left(\bar{\epsilon}\right).
\end{eqnarray}

\section{Mean-field slave-boson technique for the N-dot-S system} \label{A2}

In this Appendix we report details of the slave-boson method for the N-dot-S system.\cite{schwab99} We calculate instantaneous equilibrium Green's functions where all parameters are always taken at their time-averaged value. We therefore omit the bar and, e.g., write $\Gamma_\mathrm{N}$ instead of  $\bar{\Gamma}_\mathrm{N}$. Furthermore all Green's functions considered in this appendix are $2\times 2$ matrices in Nambu space and we omit the caret.
 
The free energy of the system can be written as a function of the Matsubara Green's function $\mathcal{G}$ [related to the retarded Green's function by the analytic continuation 
$\left( i\omega_n\rightarrow\omega+i\delta\right)$]:\cite{mahan81,ambegaokar84}
\begin{equation}\label{eq_free_energy}
F=-k_{\mathrm{B}}T \sum_{n=-\infty}^{\infty}\mathrm{Tr}\left\{\mathrm{ln} \mathcal{\tilde{G}}^{-1}\left(i\omega_n\right)\right\}+\lambda b_0^2 \ +\epsilon ,
\end{equation}
where the trace is performed in Nambu space and $\omega_n=\pi(2n+1)/\beta$ is a fermionic Matsubara frequency.
The pseudo-fermion Green's function $\mathcal{\tilde{G}}$ has the same functional form of the noninteracting Matsubara Green's function $\mathcal{G}$, but contains the renormalized parameters $\tilde{\Gamma}_{\mathrm{N}}=b_0^2\Gamma_{\mathrm{N}}$, $\tilde{\Gamma}_{\mathrm{S}}=b_0^2\Gamma_{\mathrm{S}}$ and $\tilde{\epsilon}=\epsilon+\lambda$.
Minimizing the free energy with respect to the variables $\lambda$ and $b_0$ we find the equations:
\begin{subequations}
\label{eq_slave_boson}
\begin{eqnarray}
0 & = & b_{0}^2+k_{\mathrm{B}}T\sum_{n=-\infty}^{\infty}\mathrm{Tr}\left\{\tilde{\mathcal{G}}\left(i\omega_n\right)\sigma_{z}\right\}\\
0 & = & \lambda+k_{\mathrm{B}}T\sum_{n=-\infty}^{\infty}\mathrm{Tr}\left\{\tilde{\mathcal{G}}\left(i\omega_n\right)\Gamma\left(i\omega_n\right)\right\} \ .
\end{eqnarray}
\end{subequations}
The matrix in Nambu space $\Gamma\left(i\omega_n\right)$ is given by
\begin{eqnarray}
\Gamma\left(i\omega_n\right)=
\left(\begin{array}{cc}
-i \frac{\Gamma_{\mathrm{N}}}{2}\mathrm{sign}\left(\omega_n\right)
& 0\\
0 & -i\frac{\Gamma_{\mathrm{N}}}{2}\mathrm{sign}\left(\omega_n\right)
\end{array}\right)\nonumber\\
+
\left(\begin{array}{cc}
-i\frac{\Gamma_{\mathrm{S}}}{2}\frac{\omega_n}{\sqrt{\omega_n^2+\Delta^2}}
& \frac{\Gamma_{\mathrm{S}}}{2}\frac{\Delta}{\sqrt{\omega_n^2+\Delta^2}}\\
\frac{\Gamma_{\mathrm{S}}}{2}\frac{\Delta}{\sqrt{\omega_n^2+\Delta^2}} & -i\frac{\Gamma_{\mathrm{S}}}{2}\frac{\omega_n}{\sqrt{\omega_n^2+\Delta^2}}
\end{array}\right) \ .
\end{eqnarray}
Equations (\ref{eq_slave_boson}a) and (b) are used to determine $\lambda$ and $b_0$. 
At zero temperature we can replace the sum over the Matsubara frequencies by an integral.
Furthermore, we consider the limit $\Delta\gg\Gamma_{\mathrm{S}}$ and 
we perform the following approximation \cite{schwab99}
\begin{widetext}
\begin{equation}\label{eq_approx}
\Gamma\left(i\omega_n\right)\approx
\left\{ \begin{array}{l r} 
\left(\begin{array}{cc}
-i \frac{\Gamma_{\mathrm{N}}}{2}\mathrm{sign}\left(\omega_n\right) &
\frac{\Gamma_{\mathrm{S}}}{2}\\
\frac{\Gamma_{\mathrm{S}}}{2} &
-i \frac{\Gamma_{\mathrm{N}}}{2}\mathrm{sign}\left(\omega_n\right)
\end{array}\right) & \quad\text{for } \omega<\Delta\\ \\
\left(\begin{array}{cc}
-i \left(\frac{\Gamma_{\mathrm{N}}}{2}+
\frac{\Gamma_{\mathrm{S}}}{2} \right)\mathrm{sign}\left(\omega_n\right) & 0\\
 0 & 
-i \left(\frac{\Gamma_{\mathrm{N}}}{2}+\frac{\Gamma_{\mathrm{S}}}{2}\right)\mathrm{sign}\left(\omega_n\right)
\end{array}\right) & \quad\text{for } \omega>\Delta\ .
\end{array}\right. 
\end{equation}

Under such an approximation, Eqs. (\ref{eq_slave_boson}a) and (b) read 
\begin{subequations}
\begin{eqnarray}
b_0^2 & = & 
  \frac{2\tilde{\epsilon}}
       {\pi\sqrt{\tilde{\epsilon}^2+\frac{\tilde{\Gamma}_{\mathrm{S}}^2}{4}}}
 \left[
   \arctan
       \left(
          \frac{\Delta+\frac{\tilde{\Gamma}_{\mathrm{N}}}{2}} 
               {\sqrt{\tilde{\epsilon}^2+\frac{\tilde{\Gamma}_{\mathrm{S}}^2}{4}}}
       \right)
  -\arctan
       \left(
          \frac{\frac{\tilde{\Gamma}_{\mathrm{N}}}{2}} 
               {\sqrt{\tilde{\epsilon}^2+\frac{\tilde{\Gamma}_{\mathrm{S}}^2}{4}}}
       \right)
 \right]\nonumber\\
 & + & \frac{2}
            {\pi}
 \left[
   \arctan
       \left(
          \frac{W+\frac{\tilde{\Gamma}_{\mathrm{N}}}{2}+\frac{\tilde{\Gamma}_{\mathrm{S}}}{2}} 
               {\tilde{\epsilon}}
       \right)
  -\arctan
       \left(
          \frac{\Delta+\frac{\tilde{\Gamma}_{\mathrm{N}}}{2}+\frac{\tilde{\Gamma}_{\mathrm{S}}}{2}} 
               {\tilde{\epsilon}}
       \right)
 \right]\\
\lambda & = & \frac{\Gamma_{\mathrm{N}}}{2\pi}
          \mathrm{ln}\left(
                      \frac{\left(\Delta+\frac{\tilde{\Gamma}_{\mathrm{N}}}{2}\right)^2
                             +\tilde{\epsilon}^2+\frac{\tilde{\Gamma}_{\mathrm{S}}^2}{4}}
                           {\frac{\tilde{\Gamma}_{\mathrm{N}}^2}{4}
                             +\tilde{\epsilon}^2+\frac{\tilde{\Gamma}_{\mathrm{S}}^2}{4}}
                     \right)
        +\frac{\Gamma_{\mathrm{N}}+\Gamma_{\mathrm{S}}}{2\pi} 
            \mathrm{ln}\left(
                         \frac{W^2}
                              {\left(\Delta+\frac{\tilde{\Gamma}_{\mathrm{N}}}{2}
                               +\frac{\tilde{\Gamma}_{\mathrm{S}}}{2}\right)^2+
                               \tilde{\epsilon}^2}
                       \right)\nonumber\\
 & + & \frac{\tilde{\Gamma}_{\mathrm{S}}\Gamma_{\mathrm{S}}}{2\pi\sqrt{\frac{\tilde{\Gamma}_{\mathrm{S}}^2}{4}+\tilde{\epsilon}^2}}
 \left[
   \arctan
       \left(
          \frac{\Delta+\frac{\tilde{\Gamma}_{\mathrm{N}}}{2}} 
               {\sqrt{\tilde{\epsilon}^2+\frac{\tilde{\Gamma}_{\mathrm{S}}^2}{4}}}
       \right)
  -\arctan
       \left(
          \frac{\frac{\tilde{\Gamma}_{\mathrm{N}}}{2}} 
               {\sqrt{\tilde{\epsilon}^2+\frac{\tilde{\Gamma}_{\mathrm{S}}^2}{4}}}
       \right)
 \right]\ ,
\end{eqnarray}
\end{subequations}
\end{widetext}
where we introduced a cut-off energy $W$ representing the bandwidth of the leads.

Finally, the retarded dot Green's function $G^{\mathrm{ret}}$ is obtained from
the retarded pseudo-fermion Green's function $\tilde{G}^{\mathrm{ret}}$ by means of the relation  
\begin{equation}
G^{\mathrm{ret}}=b_0^2\tilde{G}^{\mathrm{ret}} .
\end{equation}
\begin{figure}
\includegraphics[width=2.5in]{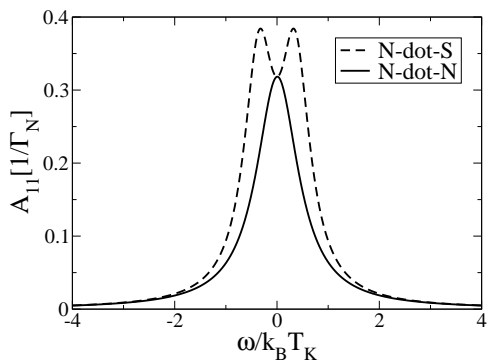}
\caption{Spectral density of the dot in the N-dot-N (solid line) and in the N-dot-S system (dashed line) in the Kondo regime ($\epsilon=-0.25 W, \ \Gamma_{\mathrm{N}}= \Gamma_{\mathrm{S}}=0.05 W$). The gap of the superconductor is half as big as the bandwidth cut-off ($\Delta=0.5 W$). The Kondo-temperature of the N-dot-N and the N-dot-S system differ by a huge factor: $T_{\mathrm{k}}^{\mathrm{N}}=7.8\times 10^{-4} W$ and $T_{\mathrm{k}}^{\mathrm{S}}=1.3\times 10^{-6} W$.
\label{fig_spectral}}
\end{figure}
From the retarded Green's function of the N-dot-S system, we can 
calculate the spectral density $A_{1,1}^{\mathrm{S}}=-(1/\pi)\mathrm{Im}G^{\mathrm{ret}}_{1,1}$ and compare to the one for the N-dot-N system, in the same regime. The result is shown in Fig. \ref{fig_spectral}. We find that the Kondo temperature in the N-dot-S system is much smaller than in the N-dot-N system (note that the width of the two curves is rescaled by the respective Kondo temperature). Nevertheless, in the deep Kondo regime, the spectral weight at the Fermi energy is the same for both systems. The strength of the Kondo-effect is therefore not affected by the presence of the superconductor in the limit of a large but finite gap $\Delta$.
\end{appendix}


\begin{thebibliography}{24}
\bibitem{note_bil} The condition for bilinear response is fulfilled if the pumped charge per infinitesimal area in parameter space depends little on the parameters, which vary within the area spanned by the pumping cycle. This condition is necessarily fulfilled if the amplitude of the time-dependent part $\delta X\left(t\right)$ is small compared to the time-average
$\bar{X}$.
\bibitem{note2} Please note that we have chosen $\Gamma_{\text{S,N}}/2$ instead of $\Gamma_{\text{S,N}}$ as pumping parameters: this choice will not be used again in the rest of the paper.
\bibitem{note1} When pumping with $\Gamma_{\mathrm{N}}$ and  $\Gamma_{\mathrm{S}}$, the pumped charge is equal to zero if the dot level is exactly at resonance (in the N-dot-S system as well as in the N-dot-N system). We therefore investigate the limit, in which the level is almost at resonance, i.e. where $\left|\epsilon\right|$ is much smaller than all other energy scales of the system.
\end{thebibliography}
\end{document}